\documentclass{article}

\usepackage{amsmath}
\usepackage{multirow}
\usepackage[ruled]{algorithm2e}
\usepackage{caption}
\usepackage{subcaption}
\usepackage{color}
\usepackage{adjustbox}

\usepackage{arxiv}

\usepackage[utf8]{inputenc} 
\usepackage[T1]{fontenc}    
\usepackage{hyperref}       
\usepackage{url}            
\usepackage{booktabs}       
\usepackage{amsfonts}       
\usepackage{nicefrac}       
\usepackage{microtype}      
\usepackage{graphicx}
\usepackage{natbib}
\usepackage{doi}
\usepackage{cleveref}       
\usepackage{autonum}  

\setcitestyle{authoryear,open={(},close={)}}

\DeclareMathOperator*{\argmin}{argmin}

\newcommand{\taubase}{\widehat{\tau}_{\text{base}}}
\newcommand{\tauim}{\widehat{\tau}_{\text{im}}}
\newcommand{\taups}{\widehat{\tau}_{\text{ps}}}

\SetKwComment{Comment}{/* }{ */}
\usepackage[normalem]{ulem}

\title{Efficient Treatment Effect Estimation with Out-of-bag Post-stratification}

\author{{Taebin Kim}\thanks{Work done while at Google LLC}\\
	University of North Carolina at Chapel Hill\\
	Chapel Hill, NC, USA\\
	\texttt{taebinkim@unc.edu}\\
	\And
	{Lili Wang}\\
	Google LLC\\
	Sunnyvale, CA, USA\\
	\texttt{wlili@google.com}\\
	\And
	{Randy Lai}\\
	Google LLC\\
	Sunnyvale, CA, USA\\
	\texttt{randylai@google.com}\\
	\And
	{Sangho Yoon}\\
	Google LLC\\
	Sunnyvale, CA, USA\\
	\texttt{shyoon@google.com}\\
}
\date{}

\begin{document}
\maketitle

\begin{abstract}
    Post-stratification is often used to estimate treatment effects with higher efficiency. However, the majority of existing post-stratification frameworks depend on prior knowledge of the distributions of covariates and assume that the units are classified into post-strata without error. We propose a novel method to determine a proper stratification rule by mapping the covariates into a post-stratification factor (PSF) using predictive regression models. Inspired by the bootstrap aggregating (bagging) method, we utilize the out-of-bag delete-D jackknife to estimate strata boundaries, strata weights, and the variance of the point estimate. Confidence intervals are constructed with these estimators to take into account the additional variability coming from uncertainty in the strata boundaries and weights. Extensive simulations show that our proposed method consistently improves the efficiency of the estimates when the regression models are predictive and tends to be more robust than the regression imputation method.

\end{abstract}


\section{Introduction}
{Controlled randomized experiments or trials are commonly used in tech and pharmaceutical companies to measure treatment effects.} Randomization balances confounding factors on different experiment arms and thus the difference in outcomes of interest between experiment arms reflect the treatment effect in an unbiased manner. 
{More precise unbiased estimators with smaller variance can reduce the required sample size and the waiting time to observe a significant difference (if one exists). Various methods are used within tech companies to reduce the variance in numerous A/B tests that they rely on. For example, Microsoft developed a single-variable adjusted method CUPED
(Controlled-experiment Using Pre-Experiment Data) for variance reduction \citep{deng2013improving}; Netflix developed a stratified sampling platform to assign users to experiments in real time \citep{xie2016improving}. At Google, multiple statistical methods were developed to increase efficiency: the Bayesian method Pre-Post that reduces the variance by conditioning on an observed pre-experiment variable (usually the same variable being measured) \citep{soriano2017percent}; methods with extra adjustment terms from different regression models \citep{hosseini2019unbiased, Amir2023optimal}; and the post-stratified method based on \citep{miratrix2013adjusting}. In this report, we focus on our recent progress in methodology development using post-stratification to reduce the variance of estimators for randomized experiments.
}

Stratified randomization has been widely used to improve the efficiency of treatment effect estimators by grouping similar subjects within each stratum \citep{kernan1999stratified}. Blocking is one popular choice of stratified randomization where units are first stratified to predefined blocks and randomized within each block. As shown in \citet{miratrix2013adjusting}, it is guaranteed to gain efficiency in estimating the population average treatment effect (PATE) using blocking, and post-stratification is asymptotically as efficient as blocking with the difference in the order of $1/N^2$. The main objective of this paper is to determine a stratification rule that provides more precise average treatment effect estimates and their variance estimates with a nominal coverage probability (e.g., $95\%$).

The gain of efficiency from post-stratification is larger when the values of potential outcomes are more homogeneous within each stratum \citep{miratrix2013adjusting}. Thus, we expect that using prognostic variables that are strongly associated with the outcomes will help determine post-stratification rules as these selected variables would be preferred in stratified randomization \citep{kernan1999stratified} and covariate adjustment \citep{kahan2014risks}. However, choosing a proper stratifying score or post-stratification factor (PSF) requires additional caution as the variance of the treatment effect estimator highly depends on the within-strata distributions of the outcomes.

To save the hassle, we utilize endogenous post-stratification \citep{abadie2018endogenous, breidt2008endogenous} to determine PSF. In particular, we train a linear regression model using control-arm data and then predict the outcomes (as the PSF) for all samples using the predictive model. Endogenous post-stratification provides a treatment effect estimator using the observed outcomes and hence the gain of efficiency is less dependent on the model assumption. Moreover, it can be shown that endogenous post-stratification provides consistent subgroup treatment effect estimators and thus the average treatment effect \citep{abadie2018endogenous}.

After obtaining the PSF from a fitted regression model, we construct strata boundaries from scratch. We implement the root cumulative density method described in \citet{dalenius1957choice} to determine strata boundaries. This allows our method to conveniently construct strata without knowing any information about the predefined categories or levels of covariates for stratification. 

In sum, we propose a novel method \emph{out-of-bag post-stratified jackknife (OPJ)} that constructs reasonable strata without losing statistical power. Drawing inspiration from the bagging (bootstrap aggregating) method \citep{breiman1996bagging, breiman1996out}, we randomly split data and aggregate the treatment effect estimates before assembling them into a final point estimate. This procedure can be viewed as an ensemble approach \citep{mendes2012ensemble, dietterich2000ensemble} for treatment effect estimation. Furthermore, we naturally infuse the resampling step with delete-D jackknife \citep{shao2012jackknife} which enables us to simultaneously compute the confidence interval of the post-stratified treatment effect estimator.

{Alternatively, there exist various other solutions which can effectively reduce the variance of estimators. However, these alternatives possess their own inherent limitations. For example, single-variable regression-model adjustments are commonly used but usually require observation of the historical data from each experiment unit \citep{deng2013improving, lin2013agnostic}. Therefore, the performance of variance reduction relies on the regression model's assumptions. Other critiques were made by \citet{freedman2008regression} that when the sample size is not sufficiently large or the regression covariates are not correlated with the outcome, the adjusted methods may not necessarily reduce the variance.
Another drawback of model training is that it requires a hold-back sample set which may harm the statistical power. Besides, the variance reduction effect largely relies on model specification if we impute predicted potential outcomes from the fitted regression models \citep{tsiatis2008covariate}.

Our proposed method provides a possible solution to overcome the aforementioned challenges by 1) providing more robust covariate-adjustment using stratification and 2) utilizing the complete sample data without a hold-back set. In spite of their differences, it should be noted that these various methods are related to each other. Post-stratification involves robust model adjustment using categorical dummy variables; the single-variable adjustment is a special case of regression-model adjustment using multiple variables. Pre-Post \citep{soriano2017percent} can be viewed as a Bayesian implementation of the same concept of adjusting for a pre-experiment variable.
}

The rest of this paper is structured as follows. \Cref{sec:treatment_effect_estimators} introduces different treatment effect estimators for comparison. \Cref{sec:endogenous_post_stratification} explains the concept of endogenous post-stratification and its capacity of variance reduction. \Cref{sec:strata_boundaries_construction} describes the root cumulative density method for constructing strata boundaries. \Cref{sec:out_of_bag_post_stratified_jackknife} then combines multiple steps into our post-stratification method. \Cref{sec:simulation_studies} presents simulation results under different scenarios and shows consistency and efficiency of the proposed method. Finally, \Cref{sec:conclusions} concludes the paper.

\section{Treatment Effect Estimators}
\label{sec:treatment_effect_estimators}
We consider the potential outcomes framework in Neyman-Rubin causal model \citep{splawa1990application, rubin1974estimating, holland1986statistics} with $N_0$ units in the control group, $N_1$ units in the treatment group, and $N=N_0 + N_1$ units in total. Assuming the stable unit treatment value assumption (SUTVA) \citep{rubin1980randomization, rubin1986comment}, let $Y_i(0)$ and $Y_i(1)$ denote unit $i$'s control potential outcome and treated potential outcome, respectively, and let $W_i$ be 1 if unit $i$ is treated and 0 if not. Then the observed outcome of unit $i$ is 
\begin{equation}
    Y_i = (1 - W_i) Y_i(0) + W_i Y_i(1).
\end{equation}
Now define the causal estimand $\tau$ as 
\begin{equation}
    \tau = g(EY(0), EY(1)),
\end{equation}
where examples of $g(\cdot)$ include $g(x, y)=y-x$ and $y/x$, which correspond to the difference and ratio of mean values of the two potential outcomes, respectively.

Under the assumption of random sampling, the baseline method (without post-stratification) estimates the mean potential outcomes 
\begin{equation}
    \overline{Y}(w) = \frac{1}{N_w} \sum_{i: W_i = w} Y_i, \quad w = 0, 1 
\end{equation}
and the treatment effect
\begin{equation}
\label{eq:baseline_estimate}
    \taubase = g(\overline{Y}(0), \overline{Y}(1)).
\end{equation}

Another method is the regression imputation method \citep{rubin1979using}. Separate linear regression models are fitted for both experiment arms to predict the control potential outcome and the treated potential outcome of each unit. Then the imputation treatment effect estimator is given by
\begin{equation}
\label{eq:imputation_estimate}
    \tauim = \frac{1}{N}\sum_{i=1}^N g(\widehat{Y}_i(0), \widehat{Y}_i(1)).
\end{equation}

In the post-stratification setting, let $K$ be the number of strata, $p_k$ be the population strata weights, and $S_i$ be the stratum label of $i$-th unit. Let $n_{k0}$, $n_{k1}$, and $n_k=n_{k0} + n_{k1}$ represent the number of control units, treated units, and total units in the $k$-th stratum, respectively. For $k=1,\ldots,K$, we denote the estimated strata weights as
\begin{equation}
    \widehat{p}_k = \frac{n_k}{N} = \frac{n_{k0} + n_{k1}}{N_0 + N_1}
\end{equation}
and the strata means as
\begin{equation}
    \overline{Y}_k(w) = \frac{1}{n_{kw}} \sum_{i: S_i = k, W_i = w} Y_i, \quad w = 0, 1.
\end{equation}
Then the post-stratified means of outcome
\begin{equation}
\label{eq:post_stratified_mean}
    \widetilde{Y}(w) = \sum_{k=1}^K \widehat{p}_k \overline{Y}_k(w), \quad w = 0, 1
\end{equation}
gives the post-stratified treatment effect estimate
\begin{equation}
\label{eq:post_stratified_estimator}
    \taups = g(\widetilde{Y}(0), \widetilde{Y}(1)).
\end{equation}

Let the variance of potential outcomes within stratum for $k = 1, \ldots, K$ be
\begin{equation}
    \sigma_k^2(w) = \frac{1}{n_k - 1} \sum_{i: S_i = k} \{Y_i(w) - \overline{Y}_k(w)\}^2, \quad w = 0, 1
\end{equation}
and the variance of potential outcomes between strata be
\begin{equation}
    \overline\sigma^2(w) = \frac{1}{N - 1} \sum_{k=1}^K n_k \{\overline{Y}_k(w) - \overline{Y}(w)\}^2, \quad w = 0, 1.
\end{equation}
\citet{miratrix2013adjusting} showed that when $\tau = EY(1) - EY(0)$ holds, the post-stratified estimator $\taups$ is more efficient than the baseline estimator $\taubase$ when the potential outcomes are similar within each stratum and different across strata. Then the difference between the variances of $\taubase$ and $\taups$ is 
\begin{align}
    Var(\taubase) - Var(\taups) = & \left\{ \frac{1}{N} \alpha_0 \overline\sigma^2(0) + \frac{1}{N} \alpha_1 \overline\sigma^2(1) \right\} \nonumber \\
    & - \left[ \frac{1}{N} \sum_{k=1}^K p_k \left\{ \alpha_{k0} \sigma_k^2(0) + \alpha_{k1} \sigma_k^2(1) \right\} \right] \label{eq:diff_in_var}
\end{align}
for some constants $\alpha_0$, $\alpha_1$, $\alpha_{k0}$, and $\alpha_{k1}$, which depend on the expected size ratios of the two arms. Equation \eqref{eq:diff_in_var} suggests that post-stratification can increase the efficiency of the estimator when the variance of potential outcomes is small within each stratum and large across strata.

\section{Endogenous Post-stratification}
\label{sec:endogenous_post_stratification}
We employ endogenous post-stratification that only uses sample data to form stratification for more efficient treatment effect estimation. The framework gives a PSF based on prediction of the control outcomes, and the formulated PSF decides the stratification rules. As suggested in \citet{kernan1999stratified}, it is desired to avoid having strata that are too small in size. We propose utilizing a mapping of the observed covariates to a scalar value (i.e., PSF) obtained by fitting a linear regression model, which uses the subjects in the control arm. The intuition behind endogenous post-stratification is that the treatment effects tend to depend on prognosis and researchers are often interested in assessing the subgroup treatment effect for subjects with higher susceptibility. For example, a chemotherapy could be more effective for patients with higher risks and an education program might further help students with lower test scores. 

Let $q$ be the number of covariates and $\boldsymbol{X}_i \in \mathbb{R}^{q}$ be the covariate vector of the $i$-th unit. We fit a linear regression model using the control -arm data and obtain the regression parameter estimates
\begin{equation}
    \widehat\beta = \left\{ \sum_{i=1}^N (1 - W_i) \boldsymbol{X}_i \boldsymbol{X}_i^T  \right\}^{-1} \left\{ \sum_{i=1}^N (1 - W_i) \boldsymbol{X}_i Y_i \right\}.
\end{equation}
Suppose that we have predetermined strata boundaries $c_0, \ldots, c_K$, then we can post-stratify the units into strata $S_1, \ldots, S_K$ where
\begin{equation}
\label{eq:endogenous_strata}
    S_k=\{ i : \widehat{Y}_i(0) = \boldsymbol{X}_i^T\widehat\beta \in (c_{k-1}, c_k] \} \quad k=1, \ldots, K.
\end{equation}
By plugging \eqref{eq:endogenous_strata} into \eqref{eq:post_stratified_mean}, \citet{abadie2018endogenous} demonstrated that subgroup treatment effect estimator for mean difference
\begin{equation}
\label{eq:subgroup_tau}
    \widehat\tau_k = \overline{Y}_k(1) - \overline{Y}_k(0)
\end{equation}
converges to a normal distribution following the central limit theorem:
\begin{equation}
    \sqrt{N} \left( \widehat\tau_k - \tau_k \right) \xrightarrow{d} N(0, V_k),
\end{equation}
where $\tau_k$ is the true subgroup treatment effect and $V_k$ is the asymptotic variance of $\sqrt{N} \widehat\tau_k$. Note that $\taups = \sum_{k=1}^K \widehat{p}_k \widehat\tau_k$ and $\tau = \sum_{k=1}^K p_k \tau_k$ for mean difference. By applying the Delta method, we have
\begin{equation}
    \sqrt{N} \left( \taups - \tau \right) \xrightarrow{d} N(0, \sum_{k=1}^K p_k^2 V_k).
\end{equation}
Therefore, $\taups$ is consistent for mean difference. Consistency of $\taups$ for mean ratio can be proved in a similar fashion using the continuous mapping theorem.

Particularly for $g(x, y) = y - x$, we  want to decrease the variance of potential outcomes within each stratum and thus the variance of $\taups$ according to equation \eqref{eq:diff_in_var}. Therefore, we expect the stratification rule in \eqref{eq:endogenous_strata} to reduce the variance of $\taups$ since it is the linear mapping of covariates that best predicts the control-arm potential outcomes, which is often predictive of the outcomes in the treatment arm. 



\section{Strata Boundaries Construction}
\label{sec:strata_boundaries_construction}

After fitting a regression model, the next step of post-stratification is strata construction. The goal is to decide a proper set of strata boundaries that provides high efficiency in estimation. Inspired by \citet{breidt2008endogenous}, we extended the root cumulative density method described in \citet{dalenius1957choice} for this step. The root cumulative density method was proposed to decide strata that minimize the variance of stratified mean outcomes under the Neyman allocation. The Neyman allocation provides the optimal strata weights in survey sampling:
\begin{equation}
    w_k = \frac{p_k \sigma_k}{\sum_{k=1}^K p_k \sigma_k}, \quad k = 1, \ldots, K,
\end{equation}
where $\sigma_k^2$ denotes the within-strata variance.

It is not guaranteed to achieve the same efficiency as using the Neyman allocation in experimental data. In fact, it is impossible to decide strata weights before deciding the stratification rule. However, \citet{cochran1977sampling} showed that the effect of deviation on variance from the weight allocation is bounded by the largest squared value of the proportional differences between the weights. This justifies the root cumulative density method since $w_k=p_k$ is satisfied under the equal within-stratum variance assumption. If the variances are not too disparate, we can use the observed strata weights as a tangible alternative of the optimal weights. Therefore, we consider the strata boundaries that minimize the variance of the stratified means under the Neyman allocation. The idea is that the optimum point under the Neyman allocation and the estimated strata weights are close to each other when the sample size is sufficiently large, i.e., 
\begin{align}
    \argmin_{w_1, \ldots, w_K} Var(\widetilde{Y}(w)) & \simeq \argmin_{\widehat{p}_1, \ldots, \widehat{p}_K} Var(\widetilde{Y}(w)), \quad w = 0, 1, \\
    \min_{w_1, \ldots, w_K} Var(\widetilde{Y}(w)) & \simeq \min_{\widehat{p}_1, \ldots, \widehat{p}_K} Var(\widetilde{Y}(w)), \quad w = 0, 1.
\end{align}

The root cumulative density method is performed as follows. Given samples, the probability density function is estimated using kernel density estimate \citep{chen2017tutorial} and denoted as $\widehat{f}$. Then the square root of $\widehat{f}$ is computed and the strata boundaries $c_0, \ldots, c_K$ are determined to satisfy
\begin{equation}
    \int_{c_0}^{c_1} \sqrt{\widehat{f}(x)} dx = \cdots = \int_{c_{K-1}}^{c_K} \sqrt{\widehat{f}(x)} dx.
\end{equation}

\begin{figure}[t]
    \centering
    \includegraphics[width=0.75\textwidth]{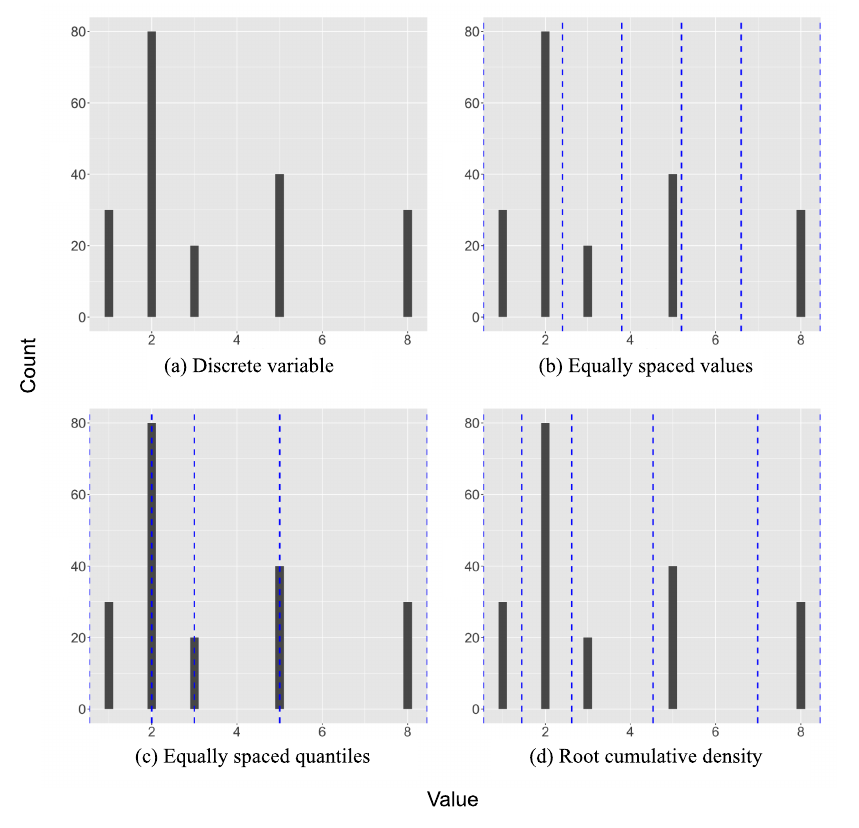}
    \caption{Example of strata boundaries construction. A discrete variable $Y$ is sampled to take the values 1, 2, 3, 5, and 8 with the corresponding counts of 30, 80, 20, 40, and 30, respectively. (a) shows the histogram of $Y$. The blue dashed lines represent the strata boundaries. Strata are constructed using the equally spaced valued method in (b), the equally spaced quantiles method in (c), and the root cumulative density method in (d). Only the root cumulative density method correctly splits the classes into separate strata.}
    \label{fig:strata_construction}
\end{figure}

\Cref{fig:strata_construction} shows an example of strata construction using different methods. In the example, a toy dataset is created by sampling a discrete variable $Y$ from the five classes 1, 2, 3, 5, and 8 where each class is sampled 30, 80, 20, 40, and 30 times, respectively. The histogram of $Y$ is shown in (a). The strata boundaries are represented by the blue dashed lines. In (b), equally spaced values are used for constructing strata. The set of equally spaced quantiles are chosen for strata boundaries in (c). (d) represents the strata boundaries determined by the root cumulative density method. We can observe that the root cumulative density method is the only strata construction method that correctly splits the true classes into separate strata.

\citet{cochran1977sampling} assumes that the units used for deciding strata boundaries are independent from the units used for estimation. One simple way of reserving independent samples for strata construction is to separate the units into the stratification group and the estimation group. Then we can use the stratification group to decide strata boundaries and use the estimation group to stratify units, estimate strata weights, and compute the treatment effect estimator $\taups$. However, this causes loss of statistical power in treatment effect estimation since the estimation group is smaller than the original data. In order to retain the power of the data, we propose out-of-bag post-stratification, which is an ensemble method that iterates sample-splitting and aggregates the estimates to compute the treatment effect estimator. The nomenclature comes from the out-of-bag idea in the bootstrap aggregating (bagging) method \citep{breiman1996out, breiman1996bagging}.

Let $M$ be the number of sample-splitting iterations, and for $m = 1, \ldots, M$, let $s_m$ be the subset of data used for deciding strata boundaries. Then the $m$-th leave-subset-out estimator $(\taups)_{(s_m)}$ is computed using the estimation data $s_m^c$ and the treatment effect estimator is computed as
\begin{equation}
\label{eq:aggregate_estimates}
    \frac{1}{M} \sum_{m=1}^M (\taups)_{(s_m)}.
\end{equation}
Rather than sampling units in $s_m$ with replacement (bootstrap) for computing \eqref{eq:aggregate_estimates}, we randomly split data (sampling without replacement) as in \cite{shao2012jackknife} to achieve a better computational efficiency.

\section{Out-of-bag Post-stratified Jackknife}
\label{sec:out_of_bag_post_stratified_jackknife}

Jackknife is a resampling method that estimates the bias and the variance of estimators \citep{miller1974jackknife}. Assume that $q \in f$, is  i.i.d. samples $X_1, \ldots, X_n \in \mathbb{R}$ and an estimator 
\begin{equation}
    \widehat\theta = T(X_1, \ldots, X_n),
\end{equation} For $i = 1, \ldots, N$, a leave-one-out point estimate
\begin{equation}
    \widehat\theta_{(i)} = T(X_1, \ldots, X_{i-1}, X_{i+1}, \ldots, X_n),
\end{equation}
is computed to define the jackknife point estimate
\begin{equation}
    \widehat\theta_{\text{jk}} = \frac{1}{N} \sum_{i=1}^N \widehat\theta_{(i)}.
\end{equation}

A variation of jackknife is group jackknife where a bucket of samples is left out for each iteration. Now, let $B$ be the number of buckets and $X_b$ be the $b$-th bucket. After grouping samples into $B$ buckets with equal proportion, a leave-bucket-out point estimate is computed as
\begin{equation}
\label{eq:group_jackknife}
    \widehat\theta_{(b)} = T(X_1, \ldots, X_{b-1}, X_{b+1}, \ldots, X_B)
\end{equation}
for $b = 1, \ldots, B$, and the jackknife point estimate is obtained as
\begin{equation}
    \widehat\theta_{\text{jk}} = \frac{1}{B} \sum_{b=1}^B \widehat\theta_{(b)}.
\end{equation}
\citet{hinkley1977jackknife} shows that we can compute the standard error of $\widehat\theta$
\begin{equation}
    \widehat{SE}(\widehat\theta) 
    = \widehat{SE}(\widehat\theta_{\text{jk}}) 
    = \sqrt{\frac{B-1}{B}\sum_{b=1}^B (\widehat\theta_{(b)} - \widehat\theta_{\text{jk}})^2}
\end{equation}
and construct a confidence interval of $\widehat\theta$ with the confidence level of $\alpha$ as
\begin{equation}
\label{eq:confidence_interval}
    \widehat\theta \pm q_{B-1}^{1-\alpha/2} \cdot \widehat{SE}(\widehat\theta)
\end{equation}
where $q_{\nu}^{t}$ is the critical value $P(T > t)$ of a random variable $T$ following the Student's t-distribution with $\nu$ degrees of freedom. In Section 3 of \citet{shao1989general}, it is shown that
\begin{equation}
    \widehat\theta_{\text{jk}} \pm q_{B-1}^{1-\alpha/2} \cdot \widehat{SE}(\widehat\theta)
\end{equation}
is asymptotically equivalent to \eqref{eq:confidence_interval} under mild regularity conditions.

\begin{figure}[t]
    \centering
    \includegraphics[width=0.9\textwidth]{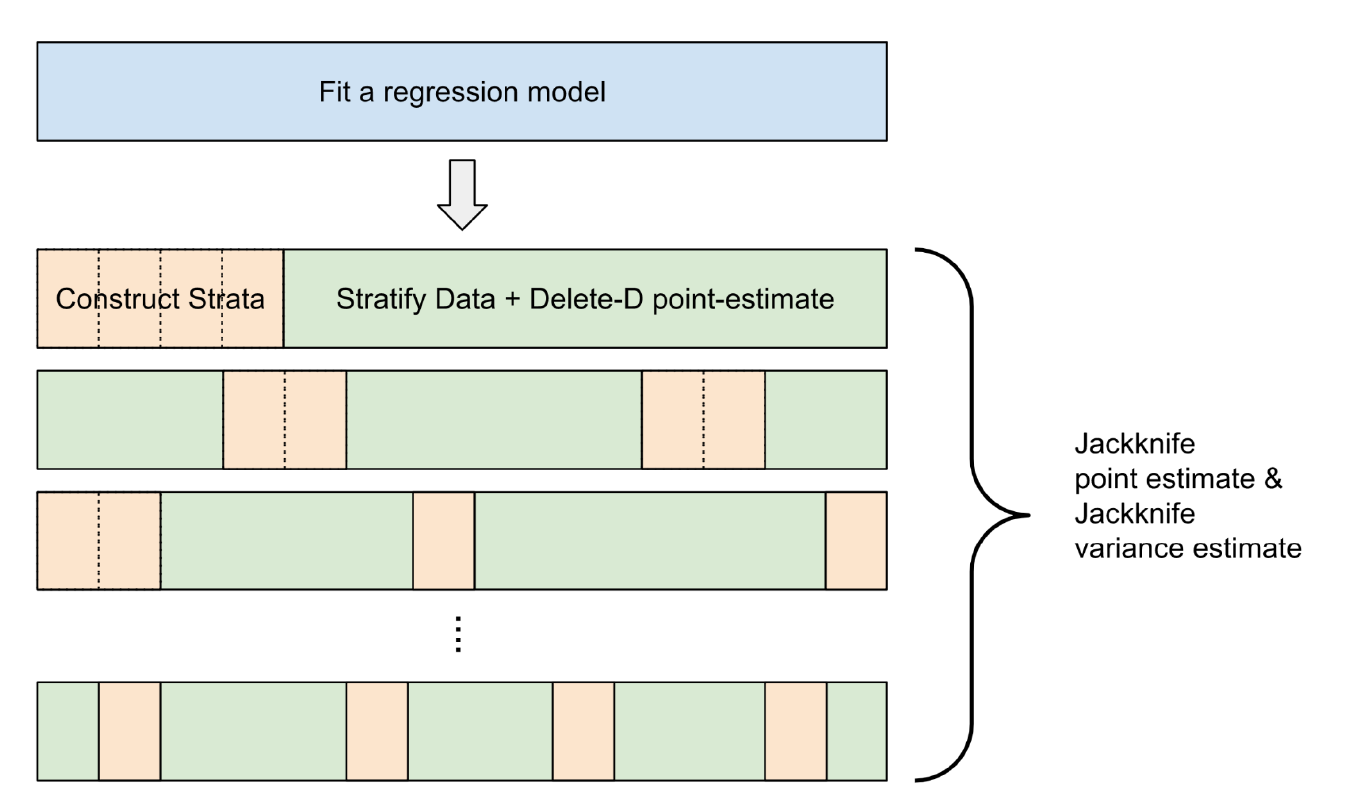}
    \caption{Diagram of out-of-bag post-stratified jackknife. All control units are used for fitting a regression model. For $m = 1, \ldots, M$, $D$ buckets are randomly chosen for strata construction, and the remaining $B-D$ buckets are used for stratifying units and computing the delete-D jackknife estimate. The $M$ jackknife estimates are aggregated to compute the jackknife point estimate and the standard error estimate of the treatment effect.}
    \label{fig:diagram}
\end{figure}

We infuse group jackknife with out-of-bag post-stratification to efficiently estimate the treatment effect and the variance of the estimator. The major restriction of ordinary group jackknife is that it limits the number of omitted bucket to one, which also limits the number of units used for strata construction. Therefore, we apply delete-D jackknife, which is when $D ~ (\geq 1)$ buckets are thrown out in each iteration, to allow flexible sample-splitting. The diagram of out-of-bag post-stratified jackknife is shown in \Cref{fig:diagram}. First, we fit a regression model using all control units in data. Then we iterate jackknife resampling for $M$ times. In each iteration, $D$ buckets are randomly selected and used for constructing strata. The remaining $B - D$ buckets are used for computing the leave-D-out estimate. 


Let $s_m$ be the collection of the deleted $D$-buckets in the $m$-th iteration. Then we compute the delete-D point estimate of the treatment effect $(\taups)_{(s_m)}$ by plugging \eqref{eq:post_stratified_estimator} into \eqref{eq:group_jackknife}, and we aggregate these values to compute the jackknife point estimate of the treatment effect $(\taups)_{\text{jk}}$ using equation \eqref{eq:aggregate_estimates}. Furthermore, it can be shown that the standard error estimate of $(\taups)_{\text{jk}}$ is
\begin{equation}
\label{eq:delete_d_standard_error}
    \widehat{SE}(\taups) = \sqrt{\frac{B-D}{DM}\sum_{m=1}^M\left( (\taups)_{(s_m)} - (\taups)_{\text{jk}} \right)^2}
\end{equation}
and that the standard error estimate in \eqref{eq:delete_d_standard_error} is consistent if $M/B \to \infty$ as $B \to \infty$ \citep{shao2012jackknife}. To save computation time, it is suggested to use $M = B^{1 + \delta}$ for a small positive value of $\delta$.



\section{Simulation Studies}
\label{sec:simulation_studies}

To evaluate the performance of our method in variance reduction, we run simulations on multiple scenarios using the baseline method without post-stratification as in \eqref{eq:baseline_estimate}, the regression imputation method as in \eqref{eq:imputation_estimate}, and the out-of-bag post-stratification method described in \Cref{sec:out_of_bag_post_stratified_jackknife}. The outcome variable $Y$, the covariates $X_1, X_2, X_3$ and the error term $\epsilon$ are generated in simulations as follows. 
\begin{align}
    X_1, X_2 & \sim N(0, 1), \\ 
    X_3 & \sim \text{Uniform}\{-\sqrt{2}, -\sqrt{2}/2, 0, \sqrt{2}/2, \sqrt{2}\}, \\
    \epsilon & \sim N(0, 1).
\end{align}
Note that $X_1$ and $X_2$ are continuous normal random variables and $X_3$ is a discrete uniform random variable.

A linear regression model is fitted using two continuous variables $X_1, X_2$ and one discrete variable $X_3$. For sanity check, we generate data where the outcome is completely independent to the covariates. To cover general cases, a scenario under perfect model assumption and a scenario under model misspecification are included in the simulations. We also compare the results of out-of-bag post-stratification to naive post-stratification, which is when a single covariate is selected from the three covariates $X_1, X_2, X_3$ as the PSF. 

In each scenario, we iterate data generation 2000 times. Throughout the simulation, the size of the control group $n_0=1000$, we fix the size of the treated group $n_1=1000$, the number of strata $K=5$, the number of buckets $B=20$, the number of deleted buckets $D=4$, and the number of resampling iteration $M=60$. An additive treatment effect is used in each case.

When the outcome is independent to the covariates, we generate the outcome as $Y = \epsilon$. In the scenario with perfect model assumption, the outcome variable $Y$ and the treatment effect are generated using linear functions of the covariates as
\begin{equation}
\label{eq:lin_lin}
    Y = 1 + 3X_1 - 2X_2 + X_3 + \frac{1}{5}(1 + X_1 + 2X_2 + 3X_3) \cdot W + \epsilon
\end{equation}
or
\begin{equation}
\label{eq:lin_const}
    Y = 1 + 3X_1 - 2X_2 + X_3 + \frac{1}{5} \cdot W + \epsilon
\end{equation}
where $W$ indicates treatment assignment. \eqref{eq:lin_lin} represents the case when the outcome and the additive effect are both linear functions of the covariates. \eqref{eq:lin_const} is when the outcome is a linear function and the additive effect is a constant functions of the covariates.

For the case where we have model misspecification, the outcome and the treatment effect are generated using quadratic functions following the equation
\begin{align}
    Y = & 1 + 3X_1 - 2X_2 + X3 + 2X_1^2 + 3X_2^2 + X_3^2\\ 
    & + \frac{1}{5} (1 + X_1 + 2X_2 + 3X_3 - X_1^2 - 2X_2 ^ 2 + 3X_3^2) \cdot W + \epsilon \label{eq:quad_quad}
\end{align}
or
\begin{equation}
\label{eq:quad_const}
    Y = 1 + 3X_1 - 2X_2 + X3 + 2X_1^2 + 3X_2^2 + X_3^2 + \frac{1}{5} \cdot W + \epsilon.
\end{equation}

Both cases have the outcome in a quadratic form. The additive effect is a quadratic function of the covariates in \eqref{eq:quad_quad} and a constant in \eqref{eq:quad_const}.

Finally, we compare the performance of our method with a naive post-stratification method using a single covariate at a time. The quadratic outcome and the quadratic additive effect in \eqref{eq:quad_quad} were used but not included in model fitting to resemble a common scenario of model misspecification.

\subsection{Independent Outcome}
\begin{table}[t]
    \caption {Comparison of treatment effect estimators under independent outcome. The estimand of interest is mean difference.}
    \label{table:independent_results}\centering
    \renewcommand{\arraystretch}{1.5}
    \begin{tabular}{c|cccc}
        \toprule
        Method & Mean bias  & Mean SE & RMSE & 95\% CI Coverage \\\toprule
        Base   & 0.000 & 0.044 & 0.045 & 0.955 \\
        Impute & 0.000 & 0.044 & 0.045 & 0.957 \\
        PS     & 0.000 & 0.044 & 0.045 & 0.951 \\\bottomrule
    \end{tabular}
\end{table}

First we ran simulations for the case when the outcome is independent to any observed covariates (not including the treatment indicator), i.e., $Y=\epsilon$, with difference in mean as the causal estimand. \Cref{table:independent_results} presents the results of the simulations. Note that all three methods are unbiased. The estimated mean standard errors and RMSE are practically identical for different estimators. The constructed 95\% confidence intervals have appropriate coverage probabilities for all methods. This shows that there is no loss of efficiency in the covariate adjustment methods even when the outcome is independent to the covariates.

\subsection{Estimation Under Different Model Assumptions}
\begin{table}[t]
    \caption {Simulation results of treatment effect estimators under different scenarios. Both mean difference and mean ratio are included in the results.}
    \label{table:results}\centering
    \renewcommand{\arraystretch}{1.5}
    \begin{tabular}{cccc|cccc}
        \toprule
        Outcome   & Additive effect & Estimand   & Method & Mean bias  & Mean SE & RMSE & 95\% CI Coverage \\\toprule
                  &                 &            & Base   & -0.002 & 0.176 & 0.180 & 0.946 \\
        Linear    & Linear          & Difference & Impute & 0.000  & 0.047 & 0.048 & 0.951 \\
                  &                 &            & PS     & -0.001 & 0.068 & 0.070 & 0.945 \\\midrule 
                  &                 &            & Base   & 0.018 & 0.202 & 0.207 & 0.947 \\
        Linear    & Linear          & Ratio      & Impute & 0.003 & 0.054 & 0.055 & 0.956 \\
                  &                 &            & PS     & 0.004 & 0.075 & 0.080 & 0.942 \\\midrule   
                  &                 &            & Base   & -0.002 & 0.172 & 0.176 & 0.947 \\
        Linear    & Constant        & Difference & Impute & 0.000  & 0.044 & 0.045 & 0.957 \\
                  &                 &            & PS     & -0.001 & 0.064 & 0.066 & 0.941 \\\midrule 
                  &                 &            & Base   & 0.019 & 0.198 & 0.204 & 0.946 \\
        Linear    & Constant        & Ratio      & Impute & 0.003 & 0.053 & 0.054 & 0.956 \\
                  &                 &            & PS     & 0.004 & 0.072 & 0.077 & 0.935 \\\midrule   
                  &                 &            & Base   & 0.003 & 0.277 & 0.284 & 0.947 \\
        Quadratic & Quadratic       & Difference & Impute & 0.005 & 0.219 & 0.225 & 0.952 \\
                  &                 &            & PS     & 0.008 & 0.199 & 0.209 & 0.945 \\\midrule 
                  &                 &            & Base   & 0.003 & 0.057 & 0.058 & 0.948 \\
        Quadratic & Quadratic       & Ratio      & Impute & 0.002 & 0.045 & 0.046 & 0.955 \\
                  &                 &            & PS     & 0.003 & 0.041 & 0.043 & 0.944 \\\midrule   
                  &                 &            & Base   & 0.003 & 0.286 & 0.293 & 0.949 \\
        Quadratic & Constant        & Difference & Impute & 0.005 & 0.233 & 0.238 & 0.951 \\
                  &                 &            & PS     & 0.009 & 0.211 & 0.220 & 0.946 \\\midrule 
                  &                 &            & Base   & 0.003 & 0.059 & 0.060 & 0.952 \\
        Quadratic & Constant        & Ratio      & Impute & 0.002 & 0.048 & 0.049 & 0.950 \\
                  &                 &            & PS     & 0.003 & 0.043 & 0.045 & 0.941 \\\bottomrule
    \end{tabular}
\end{table}

Simulation results under the model assumptions \eqref{eq:lin_lin}, \eqref{eq:lin_const}, \eqref{eq:quad_quad}, and \eqref{eq:quad_const} are presented in \Cref{table:results}. The 95\% confidence intervals show proper coverage probabilities in all cases. Both mean difference and mean ratio are included in the simulations. Note that the mean bias for the post-stratified treatment effect estimator is not significantly larger than the mean bias for the baseline treatment effect estimator in each case. This shows that the post-stratification method does not introduce bias in treatment effect estimation, which is consistent with \citet{abadie2018endogenous}.

\begin{figure}[t]
    \centering
    \includegraphics[width=\textwidth]{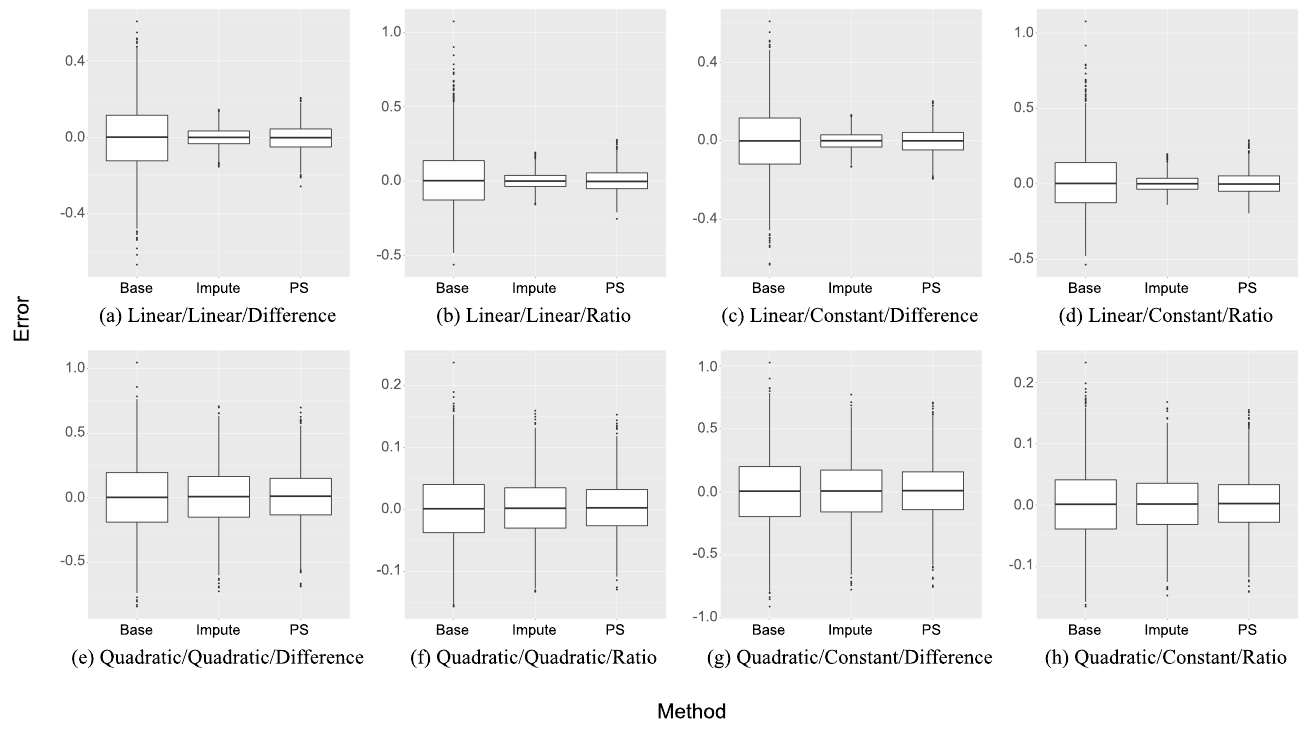}
    \caption{Boxplots of errors under different model assumptions. The subcaption summarizes the outcome, the additive effect, and the causal estimand in order. The first row represents perfect model assumption and the second represents model misspecification. The two covariate adjustment methods reduce the errors in all cases. The imputation method increases efficiency the most under perfect model assumption. The post-stratification method shows the best performance under model misspecification.}
    \label{fig:boxplots}
\end{figure}

\Cref{fig:boxplots} presents the boxplots of errors under perfect model assumption and model misspecification. The first row represents the case when the model is correctly specified, and the second row represents the case when the model is misspecified. Together with \Cref{table:results}, the mean standard error estimate, RMSE, and the distribution of the errors show that the efficiency of the treatment effect estimators are increased when one of the two covariate adjustment methods is applied in different cases. The imputation method shows the best precision when the model specification is correct. However, the post-stratification method gives the most efficient estimator when the model is misspecified. See the bottom row of \Cref{fig:boxplots} and the blocks of \Cref{table:results} where the outcome is quadratic. When the outcome is linear, the standard error estimate is reduced by up to 62\% by our method and 74\% by the imputation method. With a quadratic outcome, it is reduced by up to 28\% by our method and 21\% by the imputation method. This demonstrates that our post-stratification method is less dependent on the model assumption.



\subsection{Comparison to Naive Post-stratification}
\begin{table}[t]
    \caption {Performance of different post-stratification methods}
    \label{table:naive_results}\centering
    \renewcommand{\arraystretch}{1.5}
    \begin{tabular}{c|cccc}
        \toprule
        PSF              & Mean bias  & Mean SE & RMSE & 95\% CI Coverage \\\toprule
        $\widehat{Y}(0)$ & 0.008  & 0.199 & 0.209 & 0.945 \\
        $X_1$            & 0.008  & 0.230 & 0.234 & 0.954 \\
        $X_2$            & -0.002 & 0.236 & 0.246 & 0.945 \\
        $X_3$            & 0.003  & 0.268 & 0.278 & 0.945 \\\bottomrule
    \end{tabular}
\end{table}

We compare the proposed out-of-bag post-stratification method with naive post-stratification, which is when a PSF is selected from the covariates and the strata boundaries are constructed using equally spaced quantiles or the true classes. The model assumption in \eqref{eq:quad_quad} is used. For the continuous variables $X_1, X_2$, strata boundaries are constructed with equally spaced quantiles, and for the discrete variable $X_3$, true class separation is used for strata construction.

\Cref{table:naive_results} shows the comparison results of different post-stratification methods. The first row represents the proposed post-stratification where the PSF is determined after fitting a regression model on the control units. The second, third, and fourth rows represent naive post-stratification where the PSF is $X_1, X_2$, and $X_3$, respectively. Note that the mean bias is similar across different methods and the estimated mean standard error and RMSE show that the proposed method provides higher efficiency compared to the other naive post-stratification methods. The coverage probability of the 95\% confidence interval is in a proper range for each method.


\section{Conclusions}
\label{sec:conclusions}
Post-stratification is an effective technique that can reduce the variance of treatment effect estimators. Since stratification occurs after experiments, post-stratification is an appropriate alternative to blocking when it is not feasible. One pitfall of post-stratification is that the selection of stratification rule often depends on prior knowledge. In this paper, we propose a novel out-of-bag post-stratification method that is completely data-driven to significantly increase the efficiency of treatment effect estimates by dividing the data into proper strata using the proposed model-based PSF.

To choose a few prognostic variables for stratification, we employ endogenous post-stratification. A linear regression model is fitted using the control units and the samples are post-stratified using their predicted control outcome. This provides a PSF that is highly associated with the outcome since it is a mapping of covariate with the best linear fit. To further increase the efficiency of the post-stratified treatment effect estimator, the root cumulative density method is utilized for strata construction. In our simulation studies, it successfully separates the true levels of a discrete variable into different strata. 

\cite{abadie2018endogenous} and \cite{breidt2008endogenous} depend on the independence assumption between the samples used for strata construction and post-stratification. To guarantee the assumption, we apply out-of-bag post-stratification, which is when the dataset is repeatedly split into two parts to be separately used for each step. Infused with delete-D jackknife, the post-stratified treatment effect estimator and 
its estimated standard deviation are derived. The infused jackknife resampling takes into account the variability coming from data-driven stratification.

We compare the proposed post-stratification method with the baseline method without covariate adjustment and the imputation method. It is shown in the simulation results in \Cref{sec:simulation_studies} that all three methods are unbiased. Note that the mean predicted outcomes in the imputation method are unbiased even if the model is misspecified. When the outcome and the covariates are independent, the three methods show similar performance. In all scenarios, the proposed OPJ method and the imputation method both reduce the variance of the treatment effect estimators in comparison with the baseline method without stratification. Note that RMSE and the mean standard error estimate are similar for each case. We conclude that the proposed post-stratification method shows strong performance in variance reduction and that it is more robust to the model misspecification compared to the imputation method.

As a next step, it is desired to further automate the framework of post-stratification by providing an appropriate number of strata for users. It may be beneficial to explore other regression models such as random forest and neural networks. Further simulations can be run to better examine the behavior of out-of-bag post-stratified jackknife.



\bibliography{references}
\end{document}